\documentclass[aps,prb,twocolumn]{revtex4}
\usepackage{graphicx}
\usepackage{epsfig}
\usepackage{amsmath}
\usepackage{graphics}
\usepackage{epsfig}
\begin{document}

\title{Expectation values of  single-particle operators in the random phase approximation  ground state}

\author{D. S. Kosov }
\address{ College of Science and Engineering, James Cook University, Townsville, QLD, 4811, Australia}

\begin{abstract} 
We developed a method for computing matrix elements of single-particle operators in the correlated random phase approximation ground state.
Working with the explicit random phase approximation ground state wavefunction,  we derived practically useful and simple expression for  a  molecular property in terms of random phase approximation
amplitudes.
The theory is illustrated by the calculation of molecular dipole moments for a set of representative molecules.

\end{abstract}
\maketitle

\section{Introduction}
 Random phase approximation  (RPA) was originally  used in quantum chemistry  to describe electronically excited states of molecules. 
 Its widespread application was somewhat overshadowed by the observation that RPA does not provide a significant improvement for excited state properties in comparison to more simple configuration interaction singles.\cite{dreuw}
 Recently, RPA has been reincarnated as a  very promising method for non-perturbative correlated ground state calculations.\cite{Eshuis2012,scuseria12,scuseria15}
 Good accuracy of obtained results along with comparatively low computational cost,\cite{furche10,schurkus16}
 proper treatment of dispersion forces,\cite{dobson12} 
 possibility of the use in the context of  density functional theory\cite{scuseria15}
 and the understanding of the connection to coupled cluster method\cite{scuseria13,yang13} 
 have resulted in the significant interest in RPA theory and applications. 
 
 The main focus of RPA ground state applications has been correlation energies and other derived from total energy quantities such as reaction barriers, electron affinities, ionisation potentials.\cite{Eshuis2012,scuseria12,yang13,goerling13,rocco14,mussard15,kallay15,colonna16} 
 Simple analytical expressions  for  other molecular observables in a correlated RPA ground state, such as, for example,  electron densities and multipole moments,  are not readily available and their calculations  are  usually performed by constructing effective RPA Lagrangians.\cite{furche14,rekkedal} In this paper, we derive a general formula for  expectation value of an arbitrary single-particle operator 
  using explicit expression for RPA ground state wave-function.  Molecular observables are computed directly without use of effective Lagrangians and they are expressed in terms of Hartree-Fock matrix elements and RPA amplitudes for inverse electronic excitations.

 We have implemented
working equations within gaussian type  orbital computer program  for quantum chemical calculations\cite{mendeleev} and  illustrated the proposed theory  by the calculations of the molecular dipole moments for representative set of molecules (water, ammonia, hydrogen sulfide, hydrogen chloride, methanol and hydrogen fluoride). The results of RPA calculations are assessed  against  experiment and are also compared with M{\o}ller-Plesset second order perturbation theory (MP2)  and coupled-cluster singles and doubles method (CCSD).

\section{Theory}
\subsection{Expression for RPA correlated ground state wavefunction}
We begin with the RPA equations for separated  singlet ($J=0$) and triplet ($J=1$) electronic excitation branches
\begin{eqnarray}
\label{rpa1}
\sum_{p'h'} A^J_{ph,p'h'} X^{Ji}_{p'h'} + B^J_{ph,p'h'} Y^{Ji}_{p'h'} = \omega_{Ji} X^{Ji}_{ph}, \\
\label{rpa2}
\sum_{p'h'} B^J_{ph,p'h'} X^{Ji}_{p'h'} + A^J_{ph,p'h'} Y^{Ji}_{p'h'} = -\omega_{Ji} Y^{Ji}_{ph}.
\end{eqnarray}
Matrices $A$ and $B$ are given by the following expressions
\begin{eqnarray}
A^J_{ph,p'h'} &&= (\epsilon_p-\epsilon_h) \delta_{pp'} \delta_{hh'} 
\\
&& +\left[ 1+(-1)^J \right] (ph|p'h') -(hh'|p'p),
\nonumber
\end{eqnarray}
\begin{equation}
B^J_{ph,p'h'} =  \left[ 1+(-1)^J \right] (ph|p'h') -(p'h|ph'),
\end{equation}
where Hartree-Fock energy of $k$ molecular orbital is $\epsilon_k$  and $(kl|mn)$ is a two-electron integral in the Mulliken notations.   The RPA  excitation spectrum is $\omega_{Ji}$. Here, and throughout the paper, indices $h$ and $p$ refer to Hartree-Fock occupied and virtual molecular orbitals, respectively. 

The physical meaning of RPA amplitudes $X^{Ji}_{ph}$ and  $Y^{Ji}_{ph}$  becomes self-evident if we write explicitly the excitation creation operator (excited state with total spin $J$, spin projection $M$, and excitation energy $\omega_{Ji}$)
\begin{equation}
Q^\dagger_{JMi} = \sum_{ph} X_{ph}^{Ji} C^\dag_{ph}(JM) - Y_{ph}^{Ji} C_{ph}(\overline{JM}).
\end{equation}
Here 
\begin{equation}
C^{\dagger}_{ph}(JM) = \sum_{\sigma \sigma'} \langle\frac{1}{2} \sigma \frac{1}{2} \sigma' | JM \rangle a^\dagger_{p\sigma}  a_{\overline{h \sigma'}}
\end{equation}
creates particle-hole excited pair with  spin $J$ and spin projection $M$.
Over-line over spin indices means  angular momentum time-reversal state  $C_{ph}(\overline{JM}) = (-1)^{J+M}C_{ph}({J-M}) $ and $a_{\overline{h \sigma}}= (-1)^{1/2+\sigma} a_{h-\sigma}$. 
The Clebsch-Gordon coefficient $\langle\frac{1}{2} \sigma \frac{1}{2} \sigma' | JM \rangle$ couples electronic spins of occupied and virtual molecular orbitals into singlet $J=0, M=0$
and triplet $J=0, M=1,0,-1$ states. $a^\dag_{p\sigma} (a_{p\sigma})$  creates (annihilate) electron with spin $\sigma$ in  virtual molecular orbital $p$ and  $a^\dag_{h\sigma} (a_{h\sigma})$
 creates (annihilate) electron with spin $\sigma$ in  occupied molecular orbital $h$. 
 Within RPA approximation, operators $C^{\dagger}_{ph}(JM)$ and $C_{ph}(JM)$ behave like boson creation and annihilation operators
 \begin{equation}
 \left[ C_{ph}(JM), C^{\dagger}_{p'h'}(J'M')   \right]= \delta_{pp'} \delta_{hh'} \delta_{JJ'} \delta_{MM'}.
 \end{equation}

The excited state wavefunction is 
\begin{equation}
Q^\dagger_{JMi} |\Psi_0 \rangle. 
\end{equation}
It is built on the correlated  RPA ground state  $|\Psi_0  \rangle $, which is  different from Hartree-Fock ground state and formally defined as a vacuum for the RPA excitation annihilation operators
\begin{equation}
Q_{JMi}  |\Psi_0\rangle =0,  \text{      for all  } J, M, \text{ and } i.
\label{vacuum}
\end{equation}
 The  excitation annihilation operator is obtained from the corresponding creation operator via Hermitian conjugation
 \begin{equation}
Q_{JMi} = \sum_{ph} X_{ph}^{Ji} C_{ph}(JM) - Y_{ph}^{Ji} C^\dag_{ph}(\overline{JM}).
\end{equation}
This equation (\ref{vacuum}) can  be algebraically resolved. The solution is the RPA ground state wavefunction which is   written as exponential operator acting on Hartree-Fock ground state $|HF \rangle$ (for details of derivation we refer to\cite{rowe})
\begin{equation}
|\Psi_0 \rangle = N e^S |HF\rangle,
\label{psi0}
\end{equation}
where
\begin{equation}
S= \frac{1}{2} \sum_{JM} \sum_{php'h'} T^J_{php'h'} C^\dag_{ph}(JM)  C^\dag_{p'h'}(\overline{JM}).
\label{S}
\end{equation}
The expression for the normalisation factor $N$ is not required in our calculations. The coefficient $T$ in the exponent is obtained  from the following matrix equation
\begin{equation}
\sum_{p'h'} T^J_{php'h'} X^{Ji}_{p'h'} =  Y^{Ji}_{p'h'}.
\label{tx}
\end{equation}
The system of equations (\ref{tx})
 should be solved separately for singlet and triplet excitations, therefore the calculation of the RPA ground state correlated wavefunction requires the knowledge of entire singlet and triplet RPA spectrum.

Having obtained the explicit expression for RPA correlated ground state wavefunction, it is 
interesting to compare it directly  ring coupled cluster doubles  (ring-CCD) ground  state.  They are seemingly very similar: RPA expression (\ref{psi0}) for ground state is the same as CCD  wavefunction, and  defining coefficient $T$ through  (\ref{tx}) is equivalent to ring-CCD (compare eq.(13a) in  \cite{scuseria13} with the system of equations (\ref{tx})). There is, however, a subtle and not always fully appreciated difference between ring-CCD and RPA ground state wavefunctions.  The RPA theory treats $C^\dag_{ph}(JM) $ operators as  bosons (that generally leads to violation of the Pauli principle and double counting of the configurations included into the RPA wavefunction, see \cite{soloviev92}) whereas ring-CCD preserves the  fermionic nature of particle-hole pair creation operators properly. 

\subsection{Matrix elements of single-particle operators}
Armed with the explicit expression  for RPA ground state (\ref{psi0}), (\ref{S}), and (\ref{tx}), we are ready to compute expectation values of molecular observables.
Let us consider a general Hermitian single-particle operator
\begin{equation}
F=   \sum_\sigma \sum_{kl} f_{kl} a^\dag_{k\sigma} a_{l \sigma}.
\label{f1}
\end{equation}
where  $f_{kl}$ is the matrix element of the corresponding operator between Hartree-Fock orbitals and  $f^*_{kl}= f_{lk}$.
The central  goal in this section is the calculation of the expectation value of operator $F$ over RPA ground state wavefunction (\ref{psi0}): $\langle \Psi_0 | F|\Psi_0   \rangle$.
Main steps in the derivation  will be described in some details below. 

We first note that 
the cross-terms between virtual and occupied orbitals  $a^\dag_{p\sigma} a_{ h \sigma}$  and 
$a^\dag_{h\sigma} a_{ p \sigma}$ 
are linearly proportional to  $C^\dag_{ph}(JM)$ and $C_{ph}(JM)$,  respectively. Therefore  these terms  are also  linear in excitation creation $Q^\dag_{JMi}$ and annihilation operators $Q_{JMi}$ and do not contribute to RPA ground state expectation value  due to (\ref{vacuum}). We omit them from the expression for single-particle operator (\ref{f1}) in the beginning of the derivations. The part of single-particle operator  (\ref{f1}) that gives non-vanishing contribution to the ground state expectation value is
\begin{equation}
F=   \sum_\sigma \sum_{pp'} f_{pp'} a^\dag_{p \sigma} a_{p' \sigma} + \sum_\sigma \sum_{hh'} f_{hh'} a^\dag_{h \sigma} a_{h' \sigma}.
\label{f2}
\end{equation}
We begin with
\begin{equation}
\langle \Psi_0 | F |\Psi_0 \rangle = N \langle \Psi_0 | F e^S |HF\rangle.
\end{equation}
The commutator of $C^\dag_{ph}(JM)$ with $F$ can be readily computed
\begin{eqnarray}
&&[F, C^\dag_{ph}(JM)] 
\\
&& = \sum_{p'} f_{p' p} C^\dag_{p'h}(JM) -  \sum_{h'} f_{h h'} C^\dag_{p h'}(JM)
\nonumber
\end{eqnarray}
Using the operator identity
\begin{equation}
F e^S =
 e^S  \left(F +[F,S] + \frac{1}{2}[[F,S],S] +...  \right),
 \end{equation}
 and taking into account that  (since $[F,S] \sim C^\dag C^\dag$)
 \begin{equation}
 [[F,S],S] =0,
 \end{equation}
 we get
 \begin{eqnarray}
 \langle \Psi_0 | F |\Psi_0 \rangle = N \langle \Psi_0 |e^S  \left( F +[F,S] \right)  |HF \rangle 
 \\
 = 2 \sum_h f_{hh} + N \langle \Psi_0 |e^S [F,S]   |HF \rangle.
 \end{eqnarray}
Since $[S,F]$ commutes with $S$,    
\begin{equation}
N \langle \Psi_0 |e^S [F,S]   |HF \rangle =  \langle \Psi_0 | [F,S]   |\Psi_0 \rangle.
\end{equation}
The expectation value of single particle operator is 
 \begin{equation}
 \langle \Psi_0 | F |\Psi_0 \rangle = 2 \sum_h f_{hh} +  \langle \Psi_0 | [F,S]   |\Psi_0 \rangle .
 \end{equation}
 The commutator term can be readily computed 
 \begin{eqnarray}
 \nonumber
 \langle \Psi_0 | [F,S]   |\Psi_0 \rangle = \sum_{JM} \sum_{p_1h_1p_2 h_2} T_{p_1h_1p_2 h_2}^J 
 \\
 \left[ \sum_{p'} f_{p'p_1}  \langle \Psi_0| C^\dag_{p'h_1}(JM) C^\dag_{p_2h_2}(\overline{JM})   |\Psi_0\rangle - \right.
 \nonumber
 \\
 \left.
 - \sum_{h'} f_{h_1 h'}  \langle \Psi_0| C^\dag_{p_1h'}(JM) C^\dag_{p_2h_2}(\overline{JM})   |\Psi_0\rangle
  \right]
  \label{[fs]}
 \end{eqnarray}
 Using inverse transformation
 \begin{equation}
 C^\dag_{ph}(JM) = \sum_i  X_{ph}^{Ji} Q^\dag_{JMi} + Y_{ph}^{Ji} Q_{\overline{JM}i} 
 \end{equation}
we compute the matrix elements in ({\ref{[fs]}): 
  \begin{eqnarray}
 \nonumber
 \langle \Psi_0 | [F,S]   |\Psi_0 \rangle = \sum_{JMi} \sum_{p_1h_1p_2 h_2} T_{p_1h_1p_2 h_2}^J 
 \\
 \left[ \sum_{p'} f_{p'p_1} X_{p_2 h_2}^{Ji}  Y_{p' h_1} ^{Ji}   - \sum_{h'} f_{h_1 h'}  X_{p_2 h_2}^{Ji}  Y_{p' h_1} ^{Ji}
  \right].
  \label{[fs2]}
  \end{eqnarray}
 Contracting matrix $T$ with  RPA amplitudes $X$ by means of (\ref{tx}), we get
  \begin{eqnarray}
 \nonumber
 \langle \Psi_0 | [F,S]   |\Psi_0 \rangle = \sum_{JMi} \sum_{p_1h_1p'} \left[ f_{p'p_1} Y_{p_2 h_2}^{Ji}  Y_{p' h_1} ^{Ji}    \right.
 \\
\left. -  f_{h_1 h'}  Y_{p_2 h_2}^{Ji}  Y_{p' h_1} ^{Ji}  \right]
  \label{[fs2]}
  \end{eqnarray}

 Combining all terms together, we obtain the final expression for expectation value of single particle  matrix element
 \begin{eqnarray}
 \label{main}
&& \langle \Psi_0 | F |\Psi_0 \rangle =  2 \sum_h f_{hh}  
 \\
&& + \sum_{JMi} \sum_{p_1h_1} \left[ \sum_p f_{p p_1}  Y_{p h_1} ^{Ji} Y_{p_1 h_1}^{Ji}    -  \sum_h f_{h_1 h }  Y_{p_1 h}^{Ji}  Y_{p_1 h_1} ^{Ji}  \right].
\nonumber
 \end{eqnarray}
 The first term in (\ref{main}) is the standard Hartree-Fock value, and the second term  is resulted from post-Hartree-Fock RPA correlations in the molecular ground state.
 This expression  for a single-particle  molecular observable in the RPA ground state is one of the main results of this paper.

\section{Test Results}
All calculations have been performed with a development version of Mendeleev computer program for {\it ab initio} quantum chemical calculations \cite{mendeleev}. The current RPA implementations use Hartree-Fock  reference molecular orbitals. The RPA amplitudes  and energies are obtained using the 
following numerical scheme, which is performed separately for singlet and triplet excited states. We split matrix $\mathbf A - \mathbf B$ using Cholesky decomposition 
\begin{equation}
\mathbf A - \mathbf B =\mathbf U^T \mathbf U.
\end{equation}
The system of RPA equations (\ref{rpa1}, \ref{rpa2})  are reduced to the following symmetric eigenvalue problem
\begin{equation}
 \mathbf U (\mathbf A+ \mathbf B) \mathbf U^T \;  \mathbf Z =  \omega \;  \mathbf  Z,
 \end{equation}
 where $\omega$ is the matrix which has eigenvalues $\omega_{Ji}$ on the diagonal and zeros elsewhere.
The RPA amplitudes $X^{Ji}_{ph}$ and  $Y^{Ji}_{ph}$  are retrieved from eigenvector matrix $\mathbf Z$ by the following two step procedure:
\begin{equation}
\mathbf X +\mathbf Y = {\omega^{-1}}  \mathbf U^T \mathbf Z, 
\label{xpy}
\end{equation}
\begin{equation}
\mathbf X-\mathbf  Y=   {\omega^{-1}}  (\mathbf A+ \mathbf B)(\mathbf  X+\mathbf  Y).
\label{xmy}
\end{equation}

Having obtained RPA amplitudes from (\ref{xpy}) and (\ref{xmy}), we 
use (\ref{main})  to compute molecular dipole moments for set of representative molecules: water, ammonia, hydrogen sulfide, hydrogen chloride, methanol and hydrogen fluoride. The results of RPA calculations are  compared with  dipole moment values obtained from Hartree-Fock, MP2,   and CCSD methods.  To focus on the comparative effect of  the treatment of electronic correlations we use the same molecular geometries for all theoretical methods  and all basis sets. The molecular geometries are taken from NIST Computational Chemistry Comparison and Benchmark Database \cite{nist}. 

The results of calculations are summarised in Table I.  As is evident from Table I, Hartree-Fock based RPA performs better than MP2 and CCSD for small basis sets, when all methods generally over-spread the electron density resulting in the larger than experimental values of the dipole moments. With 6-311+G(2d,p) basis set RPA  is the best results against the experiment. As the basis set is increased further (6-311++G(2d,2p) and 6-311++G(3df,3pd)), both MP2 and CCSD shows a systematic convergence to the experimental values whereas RPA underestimate the dipole moment by 0.03-0.18 Debye. 

  \begin{widetext}
  
\begin{table}[h]
\caption{Molecular dipole moments computed within Hartree-Fock, MP2, CCSD method. Values of dipole moments are given in Debye.} 
\begin{tabular}{ l | l | c | c | c | c  | c }		
 Molecule &~~~~~~Basis set~~~~~~ & ~~~~~~HF~~~~~~ & ~~~~~~MP2~~~~~~& ~~~~~~CCSD ~~~~~~&  ~~~~~~RPA~~~~~~& Experiment\cite{nist} \\
  \hline
  \hline
 Water     & 6-31G(d)           &  2.226  &  2.185      &  2.1669     & 2.0822 &
     \\    
  &  6-31G(d,p)       & 2.1856 & 2.0999       &  2.0871     & 2.007  &
     \\    
     & 6-311+G(d,p)      &2.241    &   2.1691     &  2.1464     & 2.0178   &
     \\    
     & 6-311+G(2d,p)       &2.1627    & 2.1212        &   2.0991    &  1.9646  & 
     \\      
       & 6-311++G(2d,2p)       & 2.0618   &  1.9707        &   1.9628    &  1.8017  & 
       \\
           & 6-311++G(3df,3pd)       &2.0095    & 1.887        &   1.8913    &  1.7138  & 1.855
           \\
    \hline  
Ammonia    & 6-31G(d)            & 1.9506       &  1.9462        &1.9246     &  1.8677  &
     \\    
                    &  6-31G(d,p)        & 1.8915      &   1.8465     &  1.8249     & 1.7773  &
     \\    
                          & 6-311+G(d,p)      & 1.8329      &   1.767     &  1.7508     & 1.6882    &
     \\    
                         & 6-311+G(2d,p)      & 1.7234    &  1.6896       & 1.6754      & 1.5953   &  
                            \\      
       & 6-311++G(2d,2p)      &1.6762	&1.6147	&1.6063	&1.5163 &
       \\
           & 6-311++G(3df,3pd)       &1.6214	&1.5311 &	1.5333 &	1.4332  & 1.47

     \\        
         \hline  
 Hydrogen sulfide   & 6-31G(d)           & 1.3924     &  1.4763       &  1.4078     & 1.2897  &
     \\    
   &  6-31G(d,p)                           &  1.3736    &  1.3668    &   1.2922       &   1.2327 &
     \\    
     & 6-311+G(d,p)                       &  1.3674     &  1.3211      & 1.2478       &  1.1981  &
     \\    
     & 6-311+G(2d,p)                     &  1.2136     &   1.2220     &  1.1610     &  1.1576   & 
                              \\      
       & 6-311++G(2d,2p)      &1.1451	& 1.1051 &	1.0529 &	1.0344 &
       \\
           & 6-311++G(3df,3pd)  &   1.0599 & 	0.9808 &	0.9632 &	0.9426 &
      0.97
     \\   
       \hline  
Hydrogen chloride    & 6-31G(d)           & 1.508            &  1.5218    & 1.4651      & 1.3852  &
     \\    
   &  6-31G(d,p)                    &  1.4793         &   1.4304    & 1.3797      & 1.3328  &
     \\    
     & 6-311+G(d,p)                &   1.4447         &    1.3808    &  1.3321     & 1.2838   &
     \\    
     & 6-311+G(2d,p)             &   1.3091         &    1.2982     &   1.2545    & 1.2323   &
  \\
       & 6-311++G(2d,2p)      &1.233	& 1.1898 &	1.1535 &	1.1192 &
       \\
           & 6-311++G(3df,3pd)  &   1.1764 &	1.0964 &	1.0774 &	1.025 & 
     1.08 
          \\   
       \hline  
           Methanol    & 6-31G(d)           &  1.9399       & 1.837    & 1.8416      &  1.7354&
     \\    
      &  6-31G(d,p)                                 &  1.9146     &  1.7693    &  1.7727    &1.6781  &
     \\    
     & 6-311+G(d,p)                              &   2.0254         & 1.936     & 1.9194   & 1.7725   &
     \\    
     & 6-311+G(2d,p)                            &   1.901        &   1.8341      & 1.8205      &1.6533   & 
      \\
       & 6-311++G(2d,2p)      &1.8758	 & 1.7865 & 1.7778	& 1.5997 &
       \\
           & 6-311++G(3df,3pd)  &  1.8215	& 1.7198 & 1.8243 &	1.5245 & 
           1.7
          \\   
       \hline  
           Hydrogen flouride    & 6-31G(d)           & 1.9823   & 1.9226   & 1.91    &  1.8261&
     \\    
      &  6-31G(d,p)                                 &  1.972     &  1.8751    &  1.8687    & 1.7913 &
     \\    
     & 6-311+G(d,p)                              & 2.066           &1.9651     &1.9535    &1.836   &
     \\    
     & 6-311+G(2d,p)                            & 2.0176          &1.9415        & 1.931     &1.7983   & 
          \\
       & 6-311++G(2d,2p)      &1.9797	 & 1.8701 &	1.8686	& 1.7253 &
       \\
           & 6-311++G(3df,3pd)  &  1.9419 &	1.8196 &	1.8243 &	1.6722 & 
     1.82
     \\             
    \hline  
    \hline
  \end{tabular}
  \end{table}
\end{widetext}

\section{Conclusions}
We developed a  practical expression for values of single-particle operators in RPA correlated ground state (\ref{main}).  We  implemented
the working equations within gaussian type  orbital computer program  for quantum chemical calculations using Hartree-Fock ground state as a reference. Based on our theory, we computed RPA corrections to   values of molecular dipole moments for set of small molecules.  Our calculations show that Hartree-Fock based RPA systematically underestimate the value of the dipole moment. The Hartree-Fock based RPA theory also do not perform as good as MP2 and CCSD in computing molecular dipole moments.

The presented theory  is applicable to both Hartree-Fock and density-functional theory based RPA. It will be interesting to perform similar calculations based on Kohn-Sham orbitals to see if there is a significant improvement in comparison to Hartree-Fock based results. We have not been able to show analytically the equivalence of the proposed method and effective Largrangian approach,\cite{rekkedal,furche14}  more analytical and computational work are needed to compare both methods.


\end{document}